# Machine learning prediction of magnetic properties of Fe-based metallic glasses considering local structures

Xin Li[1,2], Guangcun Shan[1*], C.H. Shek[2]

[1]*School of Instrumentation Science and Opto-electronics Engineering, Beihang University, Beijing 100191, China*

[2]*Department of Materials Science and Engineering, City University of Hong Kong, Kowloon Tong, Hong Kong SAR, China*

[*] Corresponding author.

*E-mail address:* gshan2-c@my.cityu.edu.hk

# Abstract

Magnetism prediction is of great significance for Fe-based metallic glasses (FeMGs), which have shown great commercial value. Theories or models established based on condensed matter physics exhibit several exceptions and limited accuracy. In this work, machine learning (ML) models learned from a large amount of experimental data were trained based on eXtreme gradient boosting (XGBoost), artificial neural networks (ANN), and random forest to predict the magnetic properties of FeMGs. The XGBoost and ANN models exhibited comparably excellent predictive performance, with $R^2 \geq 0.903$, mean absolute percentage error (MAPE) $\leq 6.17$, and root mean squared error (RMSE) $\leq 0.098$. The trained ML models aggregate the influence of 13 factors, which is difficult to achieve in traditional physical models. The influence of local structure, which was represented by the experimental parameter of the supercooled liquid region, presented a significant impact on the predictive performance of ML models. The developed ML-based method here can predict the magnetic properties of FeMGs by considering multiple factors simultaneously, including complex local structures.

# 1. Introduction

Fe-based metallic glasses (FeMGs) have great commercial value for use in electrical transformers, magnetic sensors, electromagnetic wave absorbers, and magnetic amplifiers due to their excellent soft magnetic properties. These properties are characterized by high saturation magnetization, high permeability, and low coercivity. Since the first Fe-based metallic glass was developed in 1967 [1], numerous FeMGs have been investigated, mainly aiming to improve their soft magnetic properties and glass-forming ability (GFA), for example, Fe-B-Si [2–6], (Fe, Co)-B-RE [7] (RE: real earth elements), Fe-B-Mo[8–10], and Fe-B-Y [11,12]. A well-known commercial brand on the market is the FINEMET [13], which was established by Hitachi Metals.

Revealing and understanding the origin of the magnetic properties of metallic glass is of great significance for the development of new functional materials in an efficient manner. For that, some valuable theories have been proposed, such as the well-known Slater–Pauling rule [14], generalized Slater–Pauling rule [15], and theories based on band-gap or charge transfer theories [16], which are provenly feasible. However, the magnetic properties of metallic glasses are the result of many factors. These theories or models that consider insufficient factors usually have many exceptions, and their estimation accuracy is limited. Therefore, it is still a challenge to reveal an accurate variation pattern of the soft magnetic properties of FeMGs based on physiochemical theories, so a large number of experiments consider a trial-and-error strategy to develop high-performance metallic glasses

[17–26].

Recently, machine learning (ML) has been applied to material science and became a promising tool for the development of new materials with desired properties. For example, based on extensive published research data, ML models have been trained to predict the GFA of metallic glasses [27–30], mechanical properties of metallic glasses [31,32], glass transition temperature [33], hardness of high-entropy alloys [34], optical constants of 2D materials [35], and morphology of nanoscale metal–organic frameworks [36]. In the past few decades, a large number of FeMGs have been discovered, so that sufficient data have been produced, and they could meet the data needs of ML.

According to experiments and simulations, GFA and soft magnetism of FeMGs are both strongly related to their local structures [37–40]. Meanwhile, the GFA of metallic glasses can be quantified by experimental data of the supercooled liquid region ($\Delta T_x$). In this work, following the research strategy shown in Fig. 1, high-performance ML models were trained to predict the magnetic properties of FeMGs by considering local structures. A FeMGs dataset containing approximately 400 FeMGs was established based on published papers. In the FeMGs dataset, the experimental $\Delta T_x$ data represents the effect of local structures, chemical components, and other theoretical calculation parameters, and they were treated as features. The experimental data of saturated magnetization ($B_s$) were treated as labels. Three efficient ML algorithms were investigated, namely eXtreme gradient boosting (XGBoost) [41], artificial neural networks (ANN), and random forest (RF) [42]. Furthermore, the feature scores given by XGBoost indicate a strong correlation between the magnetism property and the local structure of FeMGs.

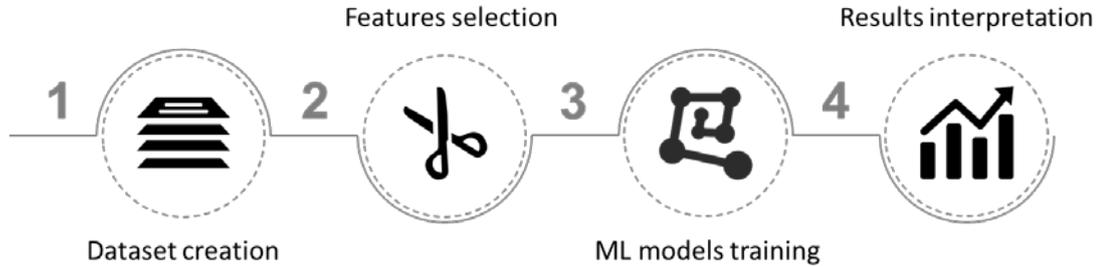

FIG. 1. Training of a machine learning model developed to reveal the relationship between soft magnetism and local structure of FeMGs.

## 2. ML methods

### 2.1 Dataset

The FeMGs dataset established in this work was obtained by searching published scientific journals that contained information about the required chemical components and experimental data, such as values of saturated magnetization ($B_s$), glass transition temperature ($T_g$), onset crystallization temperature ($T_x$), and supercooled liquid region ($\Delta T_x$). In addition, several theoretical parameters were calculated based on the chemical components of each FeMG, including theoretical density ($\rho$), theoretical melting point ($T_m$), theoretical molar volume ($V$), mean atom radius ($\bar{r}$), atomic size difference ($\delta$), configurational entropy ($\Delta S_c$), electronegativity ($\chi$), valence electron concentration ($VEC$), and valence electron concentration without FeCoNi ($VEC'$). These factors are relevant to the magnetic properties and thermal dynamics of metallic glasses. The atomic percentages of ferromagnetic single elements (Fe, Co, and Ni) and metalloids (B and Si) were also obtained. Table I details the features mentioned above. Because different features have different scales and ranges, feature normalization was conducted before training the ML models to better stabilize the training. The values of all features were normalized by

$$x' = \frac{x - \bar{x}}{\delta_x},$$

where $x$ and $x'$ refer to the original and normalized value of a feature, respectively; and $\bar{x}$ and $\delta_x$ refer to the mean and standard deviation of $x$, respectively.

Table I. Description of all features in the FeMGs dataset.

| Features | Description |
| --- | --- |
| Saturated magnetization ($B_s$) | Experimental data |
| Glass transition temperature ($T_g$) | Experimental data |
| Onset crystallization temperature ($T_x$) | Experimental data |
| Supercooled liquid region [43] ($\Delta T_x$) | $\Delta T_x = T_x - T_g$ |
| Theoretical melting point ($T_m$) | $T_m = \sum_{i=1}^{n} c_i T_{mi}$ |
| Theoretical density ($\rho$) | $\rho = 100/(\sum_{i=1}^{n} \frac{w_i}{\rho_i})$ |
| Mean atom radius ($\bar{r}$) | $\bar{r} = \sum_{i=1}^{n} c_i r_i$ |
| Theoretical molar volume ($V$) | $V = \sum_{i=1}^{n} \frac{c_i m_i}{\rho_i}$ |
| Atomic size difference [44] ($\delta$) | $\delta = \sqrt{\sum_{i=1}^{n} c_i \left(1 - \frac{r_i}{\bar{r}}\right)^2}$ |
| Configurational entropy [45] ($\Delta S_c$) | $\Delta S_c = -R \sum_{i=1}^{n} c_i \ln c_i$ |
| Electronegativity [46] ($\chi$) | $\chi = \sum_{i=1}^{n} c_i \chi_i$ |
| Valence Electron Concentration [45] (*VEC*) | $VEC = \sum_{i=1}^{n} c_i (VEC)_i$ |
| Valence Electron Concentration without FeCoNi (*VEC'*) | $VEC' = \sum_{i=1}^{n} c_i (VEC)_i - c_{Fe}(VEC)_{Fe} - c_{Co}(VEC)_{Co} - c_{Ni}(VEC)_{Ni}$ |
| Atomic ratio of Fe, Co, Ni, B, Si | $c_{Fe}, c_{Co}, c_{Ni}, c_B, c_{Si}$ |

## 2.2 Algorithm selection

From the perspective of data analysis, this work presents a nonlinear regression problem. All three algorithms studied in this work, namely XGBoost, ANN, and RF, are suitable for nonlinear regression problems with small data sizes. XGBoost is an optimized implementation of gradient boosted decision trees [41], which has shown excellent performance and fast speed in many engineering applications. ANN algorithms have been developed for a long time, and they have received significant attention after breakthroughs in hardware technology. With a feed-forward structure, ANN can learn and summarize through the experimental use of known data. Similar to XGBoost, RF is also an ensemble learning algorithm [42], which combines the prediction results of multiple base learners to obtain an optimal learner with improved generalization ability and robustness. However, XGBoost is based on serialized base learners, and RF is based on parallelized base learners.

To obtain the ML models with optimal predictive performance for the established dataset, the key parameters were tuned. For the XGBoost algorithm, the tuned parameters were *n_estimators*, *max_depth*, and *min_child_weight*. For the ANN algorithm, the tuned parameters were the number of hidden layers $n$ and number of neurons in each hidden layer $m$, which were equal in each hidden layer. For the RF algorithm, the tuned parameters were *n_estimators*.

Furthermore, the XGBoost and RF algorithms can provide feature scores, which are attributed to their intrinsic learning principles. Based on these feature scores, the relationship between $B_s$ and selected features for training can be explained. From this point of view, XGBoost and RF-based models are explainable, whereas ANN-based models are non-explainable.

## 2.3 Model training and evaluation

Machine learning is a data-driven method that requires as much data as possible. Although many published data on FeMGs were included, there is a need for caution due to possible overfitting because of insufficient data. To fully use the dataset and prevent overfitting, the *k*-fold cross-validation (*k*-CV) strategy with $k = 10$ was adopted. The dataset was randomly split into *k* partitions of the same size, which were marked as $k_i$ (i = 1, 2 … 10). The ML model was trained based on the values of $k_i$ (i = 1, 2 … 9), and the predictive performance of the model was evaluated using the values of $k_{10}$. This process was repeated *k* times to ensure that each of the *k* partitions could be treated as a test dataset. Finally, the *k* performance results were obtained, and the average value and standard deviation were used to describe the overall predictive performance of the model.

A specific regression model is defined by hyperparameters, which significantly affect the predictive performance. To quantitatively compare the predictive capabilities of different ML regression models, three evaluation metrics were calculated: determination coefficient ($R^2$), mean absolute percentage error (MAPE), and root mean squared error (RMSE).

$$R^2(y_i, \hat{y}_i) = 1 - \frac{\sum_{i=0}^{n}(y_i - \hat{y}_i)^2}{\sum_{i=0}^{n}(y_i - \bar{y})^2},$$

$$\text{MAPE}(y_i, \hat{y}_i) = \frac{1}{n}\sum_{i=1}^{n}\frac{\|y_i - \hat{y}_i\|}{\|y_i\|},$$

$$\text{RMSE}(y_i, \hat{y}_i) = \sqrt{\frac{1}{n}\sum_{i=1}^{n}(y_i - \hat{y}_i)^2},$$

where $\hat{y}_i$ is the predicted value, $y_i$ is the true value, and $\bar{y}$ is the average value of y.

The values of $R^2$ range from $-\infty$ to 1, and a value close to 1 indicates a good fit. $R^2$ is a prevalent metric that shows the variability between the predicted and true values. MAPE and RMSE are often used to compare the predictive performance of different regression models.

## 3. Results and discussions

When the size of the dataset is not sufficiently large, the number of features chosen to train the ML model should be limited to avoid overfitting and dimensionality. To investigate the correlation of the features in the original dataset, the Pearson correlation coefficient (PCC) of any two features (X, Y) can be calculated by

$$\rho_{X,Y} = \frac{E[(X-\mu_X)(X-\mu_Y)]}{\sigma_X \sigma_Y},$$

where E is the expectation, $\mu_X$ is the mean of $X$, $\mu_Y$ is the mean of $Y$, $\sigma_X$ is the standard deviation of $X$, and $\sigma_Y$ is the standard deviation of $Y$.

The PCC value ranges from -1 to 1, and an absolute value closer to 1 indicates more linearly related variables. In this work, if the PCC absolute value of two features ($|\rho_{X,Y}|$) was higher than 0.8, one of the two features was discarded based on background knowledge. Fig. 2 shows the correlation matrix generated by the PCC of each pair of features in the original dataset. The parameter pairs of $\delta$-$T_m$, $V$-$T_m$, $V$-$\rho$, $c_B$-$\delta$, $c_{Fe}$-$\Delta S_c$, and $c_B$-$V$ presented strong correlations.

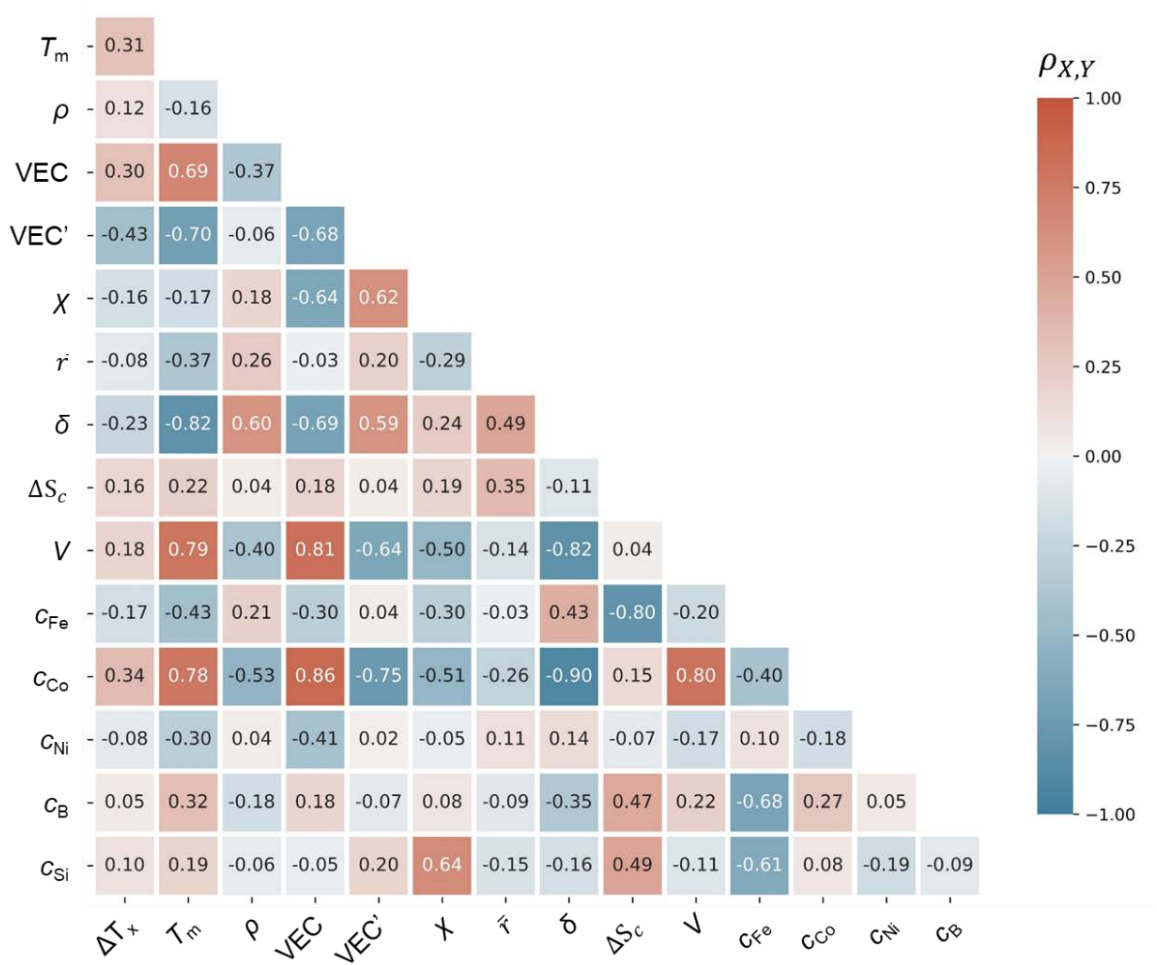

FIG. 2. Heatmap of Pearson's correlation coefficient of each pair of features in the original dataset.

Note that the k-fold cross-validation results of the three ML algorithms with 15 original features, which were denoted as XGBoost, ANN, and RF, and the three ML algorithms with reduced 13 features, which were denoted as XGBoost', ANN', and RF'. Regardless of metrics used ($R^2$, MAPE and RMSE), it was clear that the reduced features used to train the ML models had little influence on their predictive performance. The $R^2$, MAPE, and RMSE values with standard errors are listed in Table II.

Table II. Evaluation metrics for ML models.

| Metrics | XGBoost | XGBoost' | ANN | ANN' | RF | RF' |
|---|---|---|---|---|---|---|
| $R^2$ | 0.906±0.026 | 0.909±0.024 | 0.910±0.048 | 0.903±0.066 | 0.779±0.11 | 0.801±0.099 |
| MAPE (%) | 6.11±0.81 | 6.17±0.63 | 5.19±1.78 | 5.1±1.83 | 8.59±2.12 | 8.11±2.15 |
| RMSE | 0.100±0.013 | 0.098±0.010 | 0.085±0.028 | 0.085±0.033 | 0.131±0.026 | 0.124±0.024 |

According to the evaluation results, XGBoost and ANN showed comparably excellent predictive performance, whereas the predictive performance of RF was slightly inferior. It should be noted that all k-fold cross-validation results were based on tuned hyperparameters, which are discussed in detail below.

The ANN architecture used in this work is illustrated in Fig. 3(a), which contains one input layer corresponding to the 12 reduced features, $m$ hidden layers with $n$ neurons for feature extraction, and one output layer for the predicted value. The loss function used in this model is the mean squared error, which is popular for regression problems. Fig. 3(b) illustrates the effect of $m$ and $n$ values on the predictive performance of ANN, which was evaluated by RMSE. Increasing the values of $m$ and $n$ complicates the network structure. However, that did not significantly improve the model performance. The optimal hyperparameters for the ANN were $n = 64$ and $m = 2$. Fig. 3(c) shows the loss of training and validation during the learning process with a maximum epoch of 5000, which indicates that there was no overfitting in the trained ANN model.

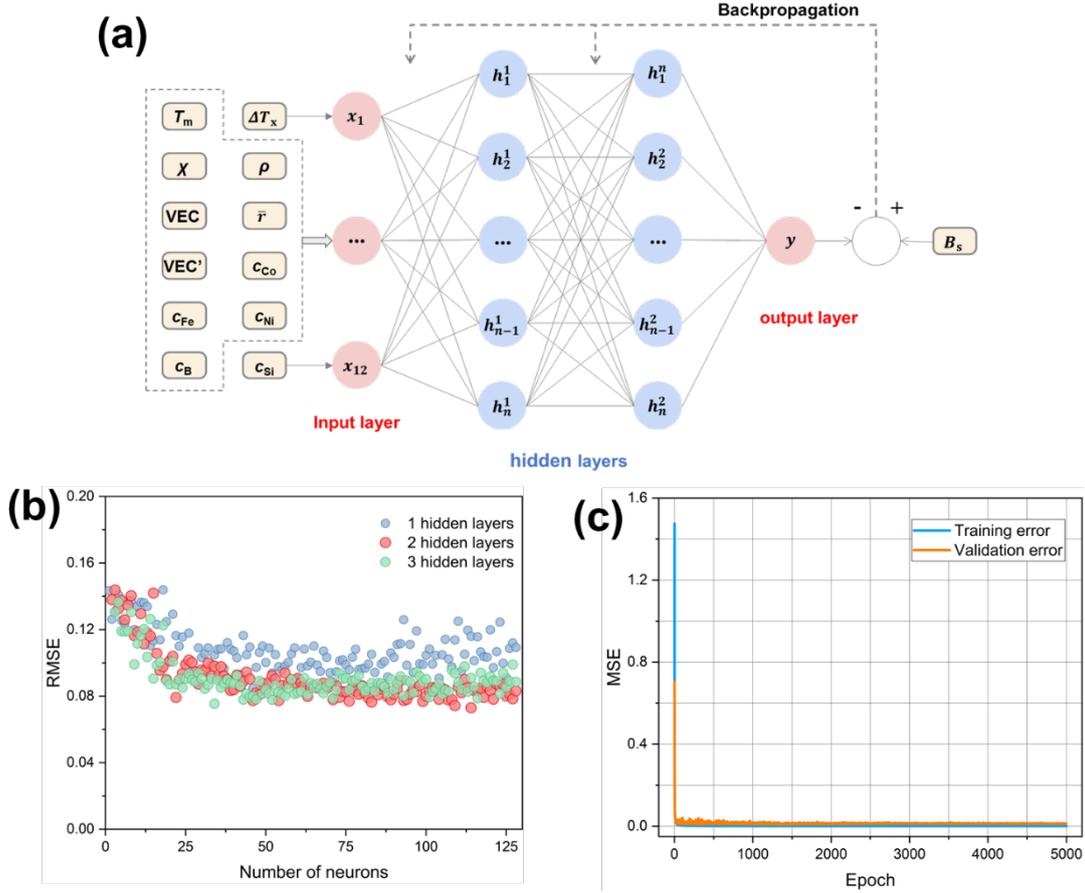

FIG. 3. (a) Schematic of the ANN architecture designed to predict $B_s$ of FeMGs based on selected features. (b) RMSE of ANN with different parameters. (c) Training and validation loss during the learning process of ANN with tuned parameters.

The hyperparameter tuning for the XGBoost models was mainly focused on *max_depth* and *min_child_weight*. Fig. 4(a) shows the heatmap of $R^2$ evaluation for the training of XGBoost models with different values of *max_depth* and *min_child_weight*. The darkest grid represents the optimal result, whose corresponding parameters are *max_depth* = 3 and *min_child_weight* = 1.

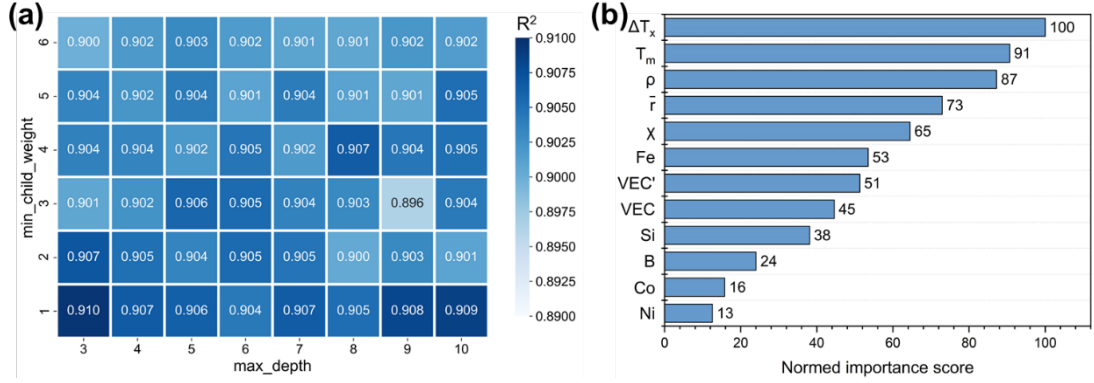

FIG. 4 (a) Hyperparameters tuning results for XGBoost evaluated by $R^2$. (b) Feature importance ranking derived from XGBoost algorithm. Importance scores were normalized by dividing the maximum score, and the maximum value was set to 100.

Based on the trained XGBoost model, the feature importance of the selected 12 features was estimated. According to the feature scores illustrated in Fig. 4(b), the supercooled liquid region ($\Delta T_x$) of FeMGs had the greatest impact on the predictive performance of the XGBoost models, followed by theoretical melting point ($T_m$) and theoretical density ($\rho$), respectively. Because $\Delta T_x$ is an experimental parameter that was used to indirectly reflect the influence of local structures, it can be concluded that the local structures of FeMGs indeed have a strong impact on their magnetic properties. Furthermore, the feature scores of $T_m$, $\rho$, $\bar{r}$, $\chi$, $VEC'$, and $c_{Fe}$ were all over 50, which indicates that the influence of other features on the predictive ability of the trained ML models cannot be ignored. In other words, the origin of the magnetic properties of FeMGs is complex, as it is influenced by multiple factors.

To simplify the theoretical analysis, the theories or models based on condensed matter physics [16,47] usually do not take into account many factors, which could be responsible for the deviation of theoretical predictions of metallic magnetism from experimental results. ML models learn from a large amount of available experimental data and consider many features simultaneously. Therefore, they are well trained to complete prediction tasks with excellent accuracy. However, the number of features should match the size of the dataset, and due to the limited dataset size, only 12 features were used to train the ML models in this work. The established theories or models provide an important reference for feature selection in ML. Furthermore, increasing the amount of data would further improve the prediction performance of the ML models.

## 4. Conclusions

In this work, based on ML models trained by XGBoost, ANN, and RF, a data-driven strategy was proposed to predict the magnetic properties ($B_s$) of FeMGs with consideration of local structures. The ML models were trained based on an FeMGs dataset, which was collected from published studies and contained nearly 400 samples. Twelve features, which quantified the FeMGs samples from many aspects, were selected to train the ML models. The hyperparameters of the three ML models were tuned using a grid search strategy. XGBoost and ANN showed comparably excellent predictive performance. XGBoost and ANN models with tuned hyperparameters also showed comparably excellent predictive performance with $R^2$ of 0.909 and 0.903, MAPE of 6.17 and 5.10, and RMSE of 0.098 and 0.085, respectively. Furthermore, a feature importance ranking was derived by XGBoost models, and it suggested that $\Delta T_x$ is crucial for the predictive performance of the ML models. In other words, the local structures of FeMGs strongly affect their magnetic properties. Compared to previous theories or models for magnetism, data-driven ML methods can aggregate the influence of many factors simultaneously, thereby providing magnetism prediction models with excellent accuracy.

# Acknowledgments

This work was financially supported by National Natural Science Foundation of China (No.21771017) and also the Fundamental Research Funds for the Central Universities.

(2017) 74–81. https://doi.org/10.1016/j.intermet.2017.01.003.